\PassOptionsToPackage{unicode}{hyperref}
\PassOptionsToPackage{hyphens}{url}
\documentclass[
]{article}
\usepackage{amsmath,amssymb}
\usepackage{lmodern}
\usepackage{iftex}
\ifPDFTeX
  \usepackage[T1]{fontenc}
  \usepackage[utf8]{inputenc}
  \usepackage{textcomp} 
\else 
  \usepackage{unicode-math}
  \defaultfontfeatures{Scale=MatchLowercase}
  \defaultfontfeatures[\rmfamily]{Ligatures=TeX,Scale=1}
\fi
\IfFileExists{upquote.sty}{\usepackage{upquote}}{}
\IfFileExists{microtype.sty}{
  \usepackage[]{microtype}
  \UseMicrotypeSet[protrusion]{basicmath} 
}{}
\makeatletter
\@ifundefined{KOMAClassName}{
  \IfFileExists{parskip.sty}{%
    \usepackage{parskip}
  }{
    \setlength{\parindent}{0pt}
    \setlength{\parskip}{6pt plus 2pt minus 1pt}}
}{
  \KOMAoptions{parskip=half}}
\makeatother
\usepackage{xcolor}
\NewDocumentCommand\citeproctext{}{}

\makeatletter
 \let\@cite@ofmt\@firstofone
 \def\@biblabel#1{}
 \def\@cite#1#2{{#1\if@tempswa , #2\fi}}
\makeatother
\newlength{\cslhangindent}
\newlength{\csllabelwidth}
\newenvironment{CSLReferences}[2] 
 {\begin{list}{}{%
  \setlength{\itemindent}{0pt}
  \setlength{\leftmargin}{0pt}
  \setlength{\parsep}{0pt}
  \ifodd #1
   \setlength{\leftmargin}{\cslhangindent}
   \setlength{\itemindent}{-1\cslhangindent}
  \fi
  \setlength{\itemsep}{#2\baselineskip}}}
 {\end{list}}
\usepackage{calc}

\usepackage{tikz}
\ifLuaTeX
  \usepackage{selnolig}  
\fi
\IfFileExists{bookmark.sty}{\usepackage{bookmark}}{\usepackage{hyperref}}
\IfFileExists{xurl.sty}{\usepackage{xurl}}{} 
\hypersetup{
  pdftitle={MoRSAIK: Sequence Motif Reactor Simulation, Analysis and Inference Kit in Python},
  hidelinks,
  pdfcreator={LaTeX via pandoc}}

\title{\texttt{MoRSAIK}: Sequence Motif Reactor Simulation, Analysis and
Inference Kit in Python}

\definecolor{c53baa1}{RGB}{83,186,161}
\definecolor{c202826}{RGB}{32,40,38}

\usepackage[affil-it]{authblk}
\usepackage{orcidlink}
\author[1,2,3,4%
  ]{Johannes Harth-Kitzerow%
    \,\orcidlink{0000-0001-5864-2258}\,%
    }
\author[3,4%
  ]{Ulrich Gerland%
    \,\orcidlink{0000-0002-0859-6422}\,%
    }
\author[1,2,4%
  ]{Torsten A. Enßlin%
    \,\orcidlink{0000-0001-5246-1624}\,%
    }

\affil[1]{Max-Planck-Institut für Astrophysik, Karl-Schwarzschild-Str.
1, 85748 Garching, Germany%
  }
\affil[2]{Ludwig-Maximilians-Universität München,
Geschwister-Scholl-Platz 1, 80539 Munich, Germany%
  }
\affil[3]{Technische Universität München, James-Franck-Str. 1, 85748
Garching, Germany%
  }
\affil[4]{Exzellenzcluster ORIGINS, Boltzmannstr. 2, 85748 Garching,
Germany%
  }
\date{01 December 2025}

\begin{document}
\maketitle

\section{Statement of need}\label{statement-of-need}

Origins of life research investigates how life could emerge from
prebiotic chemistry only. Living systems as we know them today rely on
RNA, DNA and proteins. According to the central dogma of molecular
biology, information is stored in DNA, transfered by RNA resulting in
proteins that catalyze functional reactions, such as synthesis and
replication of DNA and RNA. One possible explanation of how this
mechanism evolved provides the RNA world hypothesis (Crick 1968; Higgs
and Lehman 2014; Orgel 1968; Pressman, Blanco, and Chen 2015; Szostak
2012). It states that life could emerge from RNA strands only, storing
and transferring biological information, as well as catalyzing reactions
as ribozymes. Before this state could have emerged, however, the
prebiotic world was probably a purely chemical pool of short RNA strands
with random sequences and without biological function. Despite the lack
of guidence by proteins, the RNA sequences reacted with each other. In
such an RNA reactor RNA strands perform hybridization and
dehybridization, as well as ligation and cleavage. In this context
relevant questions are what are the conditions that allow longer RNA
strands to be built and how can information carrying in RNA sequence
emerge?

A key reaction for the emergence of longer RNA strands is templated
ligation. There, two strands hybridize adjacent onto a template strand
and ligate. The rate of this reaction is the larger, the better the two
strands match the complementary sequence of the template strand. The
extended strands can then serve as a template for the next generation of
templated ligation. This leads to an acceleration of production of
complementary strands. This process, however, is highly sensitive to
environmental conditions determining the reaction rates within an RNA
reactor (Göppel et al. 2022; Rosenberger et al. 2021).

In order to investigate those RNA reactors, efficient simulations are
needed because the space of possible RNA sequences increases
exponentially with the length of the strands, as well as the number of
reactions between two strands. In addition, simulations have to be
compared to experimental data for validation and parameter calibration.
Here, we present the \texttt{MoRSAIK} python package for sequence motif
(or k-mer) reactor simulation, analysis and inference. It enables users
to simulate RNA sequence motif dynamics in the mean field approximation
as well as to infer the reaction parameters from data with Bayesian
methods and to analyze results by computing observables and plotting.
\texttt{MoRSAIK} simulates an RNA reactor by following the reactions and
the concentrations of all strands inside up to a certain length (of four
nucleotides by default). Longer strands are followed indirectly, by
tracking the concentrations of their containing sequence motifs of that
maximum length.

\section{Summary}\label{summary}

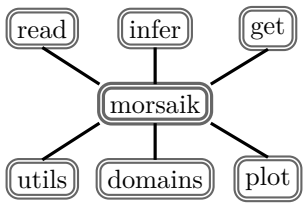
\begin{figure}
\center
\begin{tikzpicture}[
        modulenode0/.style={double,rounded corners, draw=black!60, fill=white!5, very thick, minimum size=3mm},
        modulenode1/.style={double,rounded corners, draw=black!60, fill=white!5, thick, minimum size=3mm},
        objnode1/.style={ellipse, draw=black!60, fill=white!5, very thick, minimum size=3mm},
        opnode1/.style={rectangle, draw=black!60, fill=white!5, very thick, minimum size=3mm},
    ]
    \node[modulenode0]  (morsaiknode) {morsaik};
    \node[modulenode1]  (domnode)  [below of= morsaiknode] {domains};
    \node[modulenode1]  (plotnode)  [right of= domnode, xshift=.5cm] {plot};
    \node[modulenode1]  (infernode)  [above of= morsaiknode] {infer};
    \node[modulenode1]  (readnode)  [left of= infernode, xshift=-.5cm] {read};
    \node[modulenode1]  (utilsnode)  [left of= domnode, xshift=-.5cm] {utils};
    \node[modulenode1]  (getnode)  [right of= infernode, xshift=.5cm] {get};
    \draw[-, very thick] (morsaiknode.south) -- (domnode.north);
    \draw[-, very thick] (morsaiknode.south east) -- (plotnode.north);
    \draw[-, very thick] (morsaiknode.north west) -- (readnode.south);
    \draw[-, very thick] (morsaiknode.north) -- (infernode.south);
    \draw[-, very thick] (morsaiknode.north east) -- (getnode.south);
    \draw[-, very thick] (morsaiknode.south west) -- (utilsnode.north);
\end{tikzpicture}
\caption{Overview of the MoRSAIK package modules.}
\end{figure}

The core of the \texttt{MoRSAIK}-package are chemical rate equations
simulating the reaction dynamics inside an RNA strand reactor.
\[\dot{c}_p = f_p^{(\text{ext})} + f_p^{(\text{cut})}\] The key reaction
rates are templated ligation rates, \(f_p^{(\text{ext})}\), and breakage
rates, \(f_p^{(\text{cut})}\), for the motif \(p\), where ``ext'' refers
to the extension of strands by the creation of motifs and ``cut'' for
the destruction of motifs due to cutting of strands. The terms for
templated ligation reaction rates take the form
\[f_p^{(\text{ext})} = k_\text{ext}(p,l,r,t) c_l c_r c_t,\] with the
concentrations \(c_i\) of the two reactants \(i=l\) and \(i=r\) that
form the left and right part of the product, respectively, the template
\(i=t\), and the produced motif \(i=p\). The extension rate constants
\(k_\text{ext}\) model templated ligation in a
hybridization-dehybridization equilibrium for resulting motifs or
strands \(p\). For the reactants (\(p=l\) or \(p=r\)) the same equations
apply with negative reaction rate constants. The breakage terms take the
form \[f_p^{(\text{cut})} = k_\text{cut}(p,b) c_b,\] with breaking
reactant \(i=b\) and products \(p=l,r\), as well as a negative rate for
the reactant \(p=b\). For mathematical details, please see
(Harth-Kitzerow et al. 2024) or the documentation. With \texttt{MoRSAIK}
one can simulate motif concentration trajectories given the reaction
rate constants, infer the reaction rate constants given templated
ligation counts, as well as compute the reaction rate constants based on
the free energy model of (Göppel et al. 2022) and its adaptation for
motifs (Harth-Kitzerow et al. 2024).

The \texttt{MoRSAIK}-package constists of six modules: \texttt{domains},
\texttt{read}, \texttt{infer}, \texttt{get}, \texttt{utils},
\texttt{plot}, and the objects themselves (\texttt{obj}). The
\texttt{domains}-module extends (classic) \texttt{nifty8} domains (Arras
et al. 2019) to store the \texttt{unit} (Donohue 2017) additionally to
the shape \texttt{obj}ects to ensure consistency. The
\texttt{read}-module contains a set of functions that read \texttt{yaml}
files (Simonov et al. 2024), saved arrays, and different data files such
as output and parameter files of RNA strand reactor simulations. The
\texttt{infer}-module contains all functions that compute observables
from parameters or data. The \texttt{get}-module either reads in a
stored result object if found in the result archive \texttt{MoRSAIK}
maintains or triggers the generation via the \texttt{infer}-module.

The \texttt{utils}-module is a set of different useful functions that do
not belong to one of the other modules. The \texttt{plot}-module
provides a set of plotting functions for several
\texttt{MoRSAIK}-objects. It is based on matplotlib (Hunter 2007).
\texttt{MoRSAIK}-objects are implemented in the \texttt{morsaik/obj}
directory, which is not a separate (sub)module, but imported directly
with \texttt{MoRSAIK}. Typical objects are motif vectors, which are
arrays of motif concetrations motif trajectories, time trajectories of
motifs vectors, and motif trajectory ensembles, which are ensembles of
motif trajectories. For readability, objects are designed as
\texttt{namedtuple}s on the user level. During computation they are
transformed to \texttt{Jax}-arrays (Bradbury et al. 2018) to ensure
efficient computations and differentiability, where needed. All
functions are split into small subfunctions to ensure flexibility. For
the inference, we use Geometric Variational Inference (Frank, Leike, and
Enßlin 2021; Knollmüller and Enßlin 2020) implemented in
\texttt{nifty8.re} (Edenhofer et al. 2024; Arras et al. 2019; Steininger
et al. 2019; Selig et al. 2013). For integration of ordinary
differential equations, we use \texttt{diffrax} (Kidger 2021) and
\texttt{scipy.integrate.solve\_ivp} (Virtanen et al. 2020). The
implementation in \texttt{Jax} enables fast computation and
differentiable models for inference from data with \texttt{NIFTy}.
\texttt{MoRSAIK} is the first package that provides implementation of
Bayesian inference methods for RNA reactor simulations.

Current research projects using \texttt{MoRSAIK} are the comparison of
the RNA sequence motif dynamics in an RNA strand reactor to the dynamics
in an RNA motif reactor (Harth-Kitzerow et al. 2024) and the comparison
of the infered motif dynamics to the originating strand dynamics
(Harth-Kitzerow et al. in Prep.).

\section{Acknowledgements}\label{acknowledgements}

We thank Jakob Roth, Gordian Edenhofer, Philipp Frank, Martin Reineke,
Viktoria Kainz, Tobias Göppel, Ludwig Burger, Julio C. Espinoza Campos,
Julius Lehmann and Paul Nemec for stimulating discussions. The project
was financially supported by the Deutsche Forschungsgemeinschaft (DFG,
German Research Foundation) under Germany's Excellence Strategy --
EXC-2094 -- 390783311.

\section*{References}\label{references}
\addcontentsline{toc}{section}{References}

\phantomsection\label{refs}

\end{document}